\documentclass[twocolumn,twoside,slac]{revtex4}
\usepackage{graphicx}
\usepackage{fancyhdr}
\def\nima#1#2#3  {{Nucl. Instr. and Meth. A} {\bf#1} (#2) #3.}

\begin{document}

\title{Vertex reconstruction framework and its implementation for CMS}

%

%
\author{T. Boccali}
\affiliation{CERN, Geneva, Switzerland}
\author{R. Fr\"uhwirth, W. Waltenberger}
\thanks{Supported by the Fonds zur
F\"orderung der wissen\-schaft\-lichen Forschung, Project 15177.}
\affiliation{Institut f\"ur Hochenergiephysik der \"OAW, Vienna, Austria}
\author{K. Prokofiev, T. Speer}
\affiliation{Physik-Institut der Universit\"at Z\"urich, Switzerland}
\author{P. Vanlaer}
\thanks{Corresponding author. Supported by the Belgian Federal Office 
for Scientific, Technical and Cultural affairs 
through the Interuniversity Attraction Pole P5/27.}
\affiliation{Interuniversity Institute for High Energies, ULB, Belgium}

\begin{abstract}

The class framework developed for vertex reconstruction 
in CMS is described. We emphasize how we proceed 
to develop a flexible, efficient and reliable 
piece of reconstruction software. 
We describe the decomposition of the algorithms 
into logical parts, the mathematical toolkit, 
and the way vertex reconstruction integrates into the CMS 
reconstruction project ORCA. We discuss the tools that we have 
developed for algorithm evaluation and optimization 
and for code release. 

\end{abstract}

\maketitle

\thispagestyle{fancy}


\section{INTRODUCTION}

As many reconstruction problems, vertex reconstruction 
can be decomposed into the following steps: 
\begin{itemize}
  \item Pattern recognition, or vertex finding. 
This step consists in finding clusters of compatible tracks 
among a set of tracks given on input. 
The search can either be inclusive, like in the search 
of a secondary vertex in a $b$-jet, or be guided 
by the {\it a-priori} knowledge of the decay channel. 
  \item Fitting. This step consists in 
estimating the vertex position most compatible with 
the set of tracks given on input, and constraining 
the momentum vector of the tracks using the vertex position. 
\end{itemize}

In high-luminosity experiments, vertex reconstruction algorithms 
must be able to deal with large track multiplicities, 
frequent ambiguities in the separation of primary and secondary tracks, 
and track mis-reconstructions leading to biases 
in the estimated track parameters. As an illustration 
of the difficulty to separate primary and secondary tracks, 
the resolution on the transverse impact parameter of 
the CMS tracker ranges from $\sim 100$~$\mu$m for tracks 
with $p_T = 1$~GeV/$c$ to $\sim 10$~$\mu$m for tracks 
with $p_T \ge 100$~GeV/$c$~\cite{tracker-tdr, tracking-paper}, 
while the transverse impact parameter of secondary tracks 
in $50$~GeV $b$-jets is on average much smaller than 1~mm. 

Although the algorithmic part of the problem is the most difficult to solve, 
an additional constraint comes from the CPU time available online 
in LHC experiments. Simulation studies have shown the interest of 
primary vertex finding for online event filtering~\cite{daqtdr, pvfinding}. 
In addition, $b$-jet tagging by detecting a secondary vertex 
inside the jet cone seems to perfom almost as well as 
impact parameter tagging~\cite{btag-note}, and to complement 
this method to some extent. 

For this to be applicable, the CPU time 
that vertex finding takes should be small, 
$O(50$~ms) on the processors expected in 2007. 
As vertex finding is often based on repeated 
compatibility checks between tracks and fitted 
vertices, this translates into very strong 
CPU constraints on vertex fitting algorithms 
as well.

\section{MOTIVATION FOR A FRAMEWORK}

When developing vertex reconstruction code, 
physicists face the following issues: 
\begin{itemize}
  \item The algorithmic problem is complex. The optimal 
algorithm cannot be guessed from the start, and there will 
probably be no single optimal algorithm but several, 
each optimized for a specific task. 
  \item The mathematics involved is complex, but often 
localized in a few places. Development would be made 
easier by providing a toolkit of mathematical classes. 
  \item Performance evaluation and comparison between 
algorithms is not trivial. Vertex reconstruction 
uses reconstructed tracks as input, and features 
of the vertexing algorithms must be disentangled 
from those of the tracking algorithms. Criteria 
to compare Monte Carlo simulated and reconstructed 
vertices are ill-defined and need to be standardized. 
  \item Finally, the problem is quite generic and decoupled 
from the detector geometry once tracks are given. 
This makes a good case for providing a framework 
reuseable in other HENP experiments. 
\end{itemize}

We have thus decided to develop a flexible framework 
in order to ease the development and evaluation of algorithms. 
This is realized within the CMS object-oriented reconstruction 
project ORCA~\cite{orca}. 
Section~\ref{sec:fitting} describes the vertex fitting framework. 
Section~\ref{sec:finding} deals with the vertex finding framework, 
with a focus on evaluation and optimization tools. 
Section~\ref{sec:fastsim} explains the use of a simplified 
configurable event generator for faster code development 
and release tests.

\section{VERTEX FITTING FRAMEWORK} 
\label{sec:fitting}

The fitted parameters are the solution of a minimization problem 
involving the residuals between the vertex parameters $\vec{x}$ 
and a transformation $f$ of the track parameters $\vec{p_i}$: 
$$
Min_{\vec{x}} \ F[\vec{x} - f(\vec{p_i})]; \quad i \in \text{input tracks}. 
$$
Fitting algorithms may differ by the choice of the track parametrization, 
and by the choice of the function $F$ to minimize. 
Each algorithm is a different implementation of 
the abstract {\verb VertexFitter } class. 

A usual choice for $F$ is the sum of the reduced residuals squared, 
this is the well-known Least Sum of Squares, or Least Squares, technique. 
In this case the minimum of $F$ can be expressed explicitely 
as a function of the $\vec{p_i}$'s, which is CPU-effective. 
This requires however to linearize the transformation $f$ 
in the vicinity of the true vertex position. 

\subsection{Linearization}


The linearization is performed on demand and cached for performance. 
This is done by instances of a concrete class, the {\verb LinearizedTrack }. 
Currently we use 2 linearizations, corresponding to different 
track approximations:
\begin{itemize}
  \item A straight line approximation, and a constant track 
error matrix hypothesis. 
  \item A helix approximation, and a linear error propagation 
using the first derivatives of $f$ with respect to 
the track parameters as Jacobians. 
\end{itemize}
Formally the second approximation is much more precise, 
but in the $p_T$ range of interest at the LHC, both have negligible 
contributions to the precision of the fitted vertex. 

The {\verb LinearizedTrack } is also responsible for providing 
the parametrization required by the algorithm. All useful 
parametrizations are supported. We currently use $(x, y, z)$ 
at the point of closest approach to the vertex, 
together with the straight line approximation, 
and $(q/p_T, \theta, \phi_p, d_0, z_p)$ the 5 track parameters 
at the perigee, with the helix approximation. 


Linearization around the true vertex position requires 
a first guess of this position. This is provided 
by {\verb LinearizationPointFinder } algorithms. 
These compute the average of the crossing points 
of $N$ track pairs ($N = 5$ by default). In order to use tracks 
of best precision, the $2*N$ tracks of highest $p_T$ are selected. 
Two implementations are available, one using the arithmetic average 
of the crossing points, and one using a Least Median of Squares (LMS) 
robust averaging technique~\cite{moscow, wolfgang-chep2003}. 

These algorithms rely on fast computation of the points of closest 
approach of 2 tracks. The system of 2 transcendent equations is solved 
for the running coordinates along the tracks using 
a highly optimized Newton iterative method. 
The maximum CPU time required for finding the linearization point 
is 0.1~ms on 1~GHz processors. 

\subsection{Iterative vertex fitting}

Iterations arise naturally when: 
\begin{itemize}
  \item The linearization point is too far from the fitted vertex. 
  \item The function $F$ has no explicit minimum. 
This is the case in robust fitting techniques. 
Robust estimators can be reformulated as iterative, re-weighted 
Least Squares estimators~\cite{moscow, wolfgang-chep2003}. 
  \item Several vertices are fitted together, accounting 
for ambiguities in an optimal way~\cite{multitrackfit}. 
In such a Multi-Vertex Fit, one {\verb LinearizedTrack } 
can contribute to several vertices with different weights. 
\end{itemize}

This lead us to the introduction of another concrete component, 
the {\verb VertexTrack }. It relates a {\verb LinearizedTrack } 
to a vertex, and stores the weight with which the track contributes 
to the fit. To avoid having to care for the ownership of these 
objects, {\verb LinearizedTrack } and {\verb VertexTrack } 
are handled through reference-counting pointers. 

\subsection{Sequential vertex update}

Apart from cases where some hits are shared, tracks are uncorrelated. 
This allows sequential update of the fitted parameters, 
by adding one track at a time. This procedure is faster 
than a global fit~\cite{sequentialfit}. 
The {\verb VertexUpdator } is the component responsible for updating 
the fitted parameters with the information of 1 track. 

To compute the $\chi^2$ increment, the {\verb VertexUpdator } 
uses the {\verb VertexTrackCompatibilityEstimator }. 
This increment can also be used 
at vertex finding to test the compatibility between a track 
and a vertex. The {\verb VertexUpdator } and 
the {\verb VertexTrackCompatibilityEstimator } are abstract 
and there are 2 implementations, 1 for each parametrization. 

The CPU time for track linearization and vertex update 
is $< 0.25$~ms per track on 1~GHz processors. 

\subsection{Constraining the track momentum vectors}

The momentum vectors of the tracks can also be improved 
using the vertex as a constraint. 
This is done by considering the $3 N_{tracks}$ momentum components 
as additional parameters in the vertex fit. 
However the calculation of the constrained momenta 
and their correlations is CPU-consuming, and often only 
the fitted vertex position is needed. 

We could separate this step in a {\verb VertexSmoother } component. 
It is run after fitting the vertex coordinates, using 
intermediate results cached into the {\verb VertexTrack } objects. 
The user configures the {\verb VertexFitter } so as to use 
this component or not.

\section{VERTEX FINDING FRAMEWORK}
\label{sec:finding}

The variety of vertex finding algorithms that we currently 
explore is large~\cite{wolfgang-chep2003}, and 
the decomposition of algorithms into components 
still evolves while new common features appear. 
We will thus not describe them, but rather focus 
on the evaluation and optimization tools. 

\subsection{Evaluation}

\subsubsection{Performance estimation}

We wish to compare the performance of the algorithms 
in finding the primary and secondary vertices, 
in producing ghost vertices from random track associations, 
and in assigning the tracks to their vertex efficiently 
and with a good purity. 
The first 3 figures concern the vertex finding proper, 
and are evaluated by implementations 
of the {\verb VertexRecoPerformanceEstimator } class. 
The last 2 figures concern the assignment of tracks to vertices. 
They are computed by implementations 
of the {\verb VertexTrackAssignmentPerformanceEstimator }. 
These estimators are updated each event. 

\subsubsection{Vertex selection and association}

The user needs to tell to these estimators 
which simulated vertices are important to reconstruct. 
A standard {\verb Filter } is provided, 
which keeps only the simulated vertices for which 
at least 2 tracks were successfully reconstructed. 
This allows to study the algorithmic efficiency of vertex finding. 

The user also tells how to associate a simulated vertex 
to a reconstructed one. This is defined by a {\verb VertexAssociator }. 
Association can be done by distance or by tracks, counting 
the tracks common to the simulated and the reconstructed vertex. 
In standard tests the association is done by tracks. 
A simulated vertex is considered found if there is 
a reconstructed vertex with $>50\%$ of its tracks 
originating from it. The association of the reconstructed 
and simulated tracks is performed by a {\verb TrackAssociator } 
provided by the ORCA track analysis framework. 

\subsection{Optimization}

Vertex finding algorithms have a few parameters 
that need to be tuned in order to get optimal performance 
for a given data set. Often, every physicist has to write 
his own code which scans the parameter range and finds 
the optimum parameter values. 

We propose a simple and elegant automatic tuning framework. 
The components of this framework are an abstract 
{\verb TunableVertexReconstructor }, which interacts with 
a {\verb FineTuner } object, and a special run controller 
which stops analysing events when the desired precision 
on the tuned parameters is reached. 
The {\verb TunableVertexReconstructor } provides 
the {\verb FineTuner } with the initial range of parameter values 
to be scanned. The {\verb FineTuner } is an interface 
to an optimization engine. It maximises a {\verb Score }, 
which is a function of the performance estimators 
configurable by the user: 
$$
Score = (Eff PV)^a . (Eff SV)^b . (1-Fake Rate)^c...
$$
The user only has to re-implement the {\verb TunableVertexReconstructor } 
for the concrete algorithm and parameters that are to be tuned. 
Currently the {\verb FineTuner } implementation available 
is a 1D maximizer, allowing to tune 1 parameter at a time.

\section{TESTS WITH CONTROLLED INPUT}
\label{sec:fastsim}

Parts of vertex reconstruction algorithms rely on models 
describing prior information. This information is for example 
the event topology (size of beam spot, number of vertices,...) 
or the track parameter distributions (Gaussian resolution 
and pulls, tails...). Data often depart from these models, 
which affects the performance of the algorithms. 

To disentangle this from more intrinsic features of the algorithms, 
we have developed a simple Monte Carlo generator. 
It allows to simulate data in perfect agreement 
with the prior assumptions, or distorted in a controlled way. 
The number and position of the vertices, the number and momentum 
of the decay prongs, the track parameter resolutions and tails 
are defined by the user. 

This simulation takes $O(10)$~ms per event, more than a thousand times 
faster than full event reconstruction. The time needed to perform 
most of the code debugging is reduced by the same factor. 
Another important advantage is that most of the release tests 
of the vertex reconstruction code 
can be run independently of the status of the reconstruction chain 
upstream. The simple Monte Carlo generator is currently being 
interfaced with the CMS fast Monte Carlo simulation (FAMOS)~\cite{famos} 
to account for track reconstruction effects in a more realistic way.

\section{CONCLUSION}

We have developed an efficient and flexible class framework 
for the development of vertex fitting algorithms. 
It allows the coding of usual least-squares algorithms 
as well as robust ones. 
We provide a friendly environment for the evaluation 
and optimization of vertex finding algorithms. 
As shown at this conference~\cite{wolfgang-chep2003}, 
the performance of the vertex fitters and finders tested 
for the CMS experiment are already close to matching 
the requirements for use online at the LHC.


\begin{thebibliography}{9}   


\bibitem{tracker-tdr} CMS Collaboration, 
{\it The Tracker Project - Technical Design Report}, CERN/LHCC 98-6, 1998. 

\bibitem{tracking-paper} A. Khanov et al., {\it Tracking in CMS: 
software framework and tracker performance}, \nima{478}{2002}{460} 

\bibitem{daqtdr} CMS Collaboration, {\it The TriDAS project - 
Technical Design Report, Volume 2: Data acquisition and High Level Trigger}, 
CERN/LHCC 02-26, 2002. 

\bibitem{pvfinding} D. Kotlinski, A. Nikitenko and R. Kinnunen, 
{\it Study of a Level-3 Tau Trigger with the Pixel Detector}, 
CMS Note 2001/017. 

\bibitem{btag-note} G. Segneri, F. Palla, 
{\it Lifetime based $b$-tagging with CMS}, CMS Note 2002/046. 

\bibitem{orca} http://cmsdoc.cern.ch/orca/

\bibitem{moscow} R. Fr\"uhwirth et al., {\it New developments in vertex 
reconstruction for CMS}, \nima{502}{2003}{699} 

\bibitem{wolfgang-chep2003} see W. Waltenberger, 
{\it Vertex reconstruction algorithms in CMS}, these proceedings. 

\bibitem{multitrackfit} R. Fr\"uhwirth, A. Strandlie, 
{\it Adaptive multi-track fitting}, 
Computer Physics Communications 140 (2001) 18. 

\bibitem{sequentialfit} R. Fr\"uhwirth et al., {\it Vertex reconstruction 
and track bundling at the LEP collider using robust algorithms}, 
Computer Physics Communications 96 (1996) 189. 

\bibitem{famos} S. Wynhoff, {\it Dynamically Configurable System 
for Fast Simulation and Reconstruction for CMS}, these proceedings. 
http://cmsdoc.cern.ch/famos/


\end{thebibliography}

\end{document}